\documentclass[twocolumn,superscriptaddress,showpacs,preprintnumbers,amsmath,amssymb,prl]{revtex4}
\usepackage{graphicx,bm}
\usepackage[dvips]{color}              
\newcommand{\caco}{$\rm Ca_3Co_2O_6$}

\definecolor{magenta}{rgb}{.5,0,.5}

\begin{document}
\title{Slow Magnetic Order-Order Transition in the Spin Chain Antiferromagnet Ca$_3$Co$_2$O$_6$}

\author{S.~Agrestini}
  \affiliation{Laboratoire CRISMAT, CNRS UMR 6508, ENSICAEN - Boulevard du Mar\'{e}chal Juin, 14050 Caen, France}
  \affiliation{Max Planck Institute for Chemical Physics of Solids, N\"othnitzerstr. 40, 01187 Dresden, Germany}

\author{C.~L.~Fleck}
  \affiliation{Department of Physics, University of Warwick, Coventry, CV4 7AL, UK}

\author{L.~C.~Chapon}
  \affiliation{ISIS facility, Rutherford Appleton Laboratory, Didcot OX11 0QX, UK}

\author{C.~Mazzoli}
  \affiliation{ European Synchrotron Radiation Facility, BP 220, 38043 Grenoble Cedex 9, France}

\author{A. Bombardi}
  \affiliation{Diamond Light Source Ltd., Rutherford Appleton Laboratory, Didcot OX11 0DE, UK }

\author{M.~R.~Lees}
  \affiliation{Department of Physics, University of Warwick, Coventry, CV4 7AL, UK}

\author{O.~A.~Petrenko}
  \affiliation{Department of Physics, University of Warwick, Coventry, CV4 7AL, UK}

\date{\today}

\begin{abstract}
Using powder neutron diffraction we have discovered an unusual magnetic order-order transition in the Ising spin chain compound Ca$_3$Co$_2$O$_6$. On lowering the temperature an antiferromagnetic phase with a propagation vector $\textbf{k}=\left(0.5,-0.5,1\right)$ emerges from a higher temperature spin density wave structure with $\textbf{k}=\left(0, 0, 1.01\right)$. This transition occurs over an unprecedented time-scale of several hours and is never complete.
\end{abstract}

\maketitle
Time-dependent phenomena play a crucial role in the fundamental properties of many magnetic systems.
Time dependence can arise as a result of frustration or a degeneracy in the magnetic ground state.
In model spin glasses, site or bond randomness, coupled with competing magnetic interactions, lead to cooperative spin-freezing
with a wide distribution of relaxation times. Significant time dependence (relaxation) is also seen in the isothermal and thermal remanent magnetization of spin glasses~\cite{Mydosh}.
A slow relaxation of the magnetization is observed in some single molecule magnets where the application of a magnetic field lifts the degeneracy of the ground state.
When this field is removed, the system relaxes back into equilibrium via a coupling of the spins to the environment~\cite{Gatteschi03}.
Time-dependent behavior can also be observed in superparamagnetic systems such as single domain ferromagnetic (FM) particles, in which the moments appear to freeze at some blocking temperature, as the fluctuations of the moments become slower than the time window of the probe~\cite{Bedanta}. In many of these processes the time dependence of some order parameter $P(t)$ can be modeled using a stretched exponential with $P\left(t\right)=P_{eq}+\left[P_0-P_{eq}\right]\exp\left[-\left(t/\tau\right)^\beta\right]$ and $0\leq\beta\leq1$, where $P_0$ and $P_{eq}$ are the initial and equilibrium values of $P$ respectively, and $\tau$ is a characteristic time for the process.
If the process involves overcoming an energy barrier $\Delta$ with an attempt frequency $\tau_0^{-1}$, it can follow an Arrhenius law where $\tau$ in the limit $k_BT \ll \Delta$ is $\tau=\tau_0e^{\frac{\Delta}{k_BT}}$.
The measured value of $\tau$ can vary with temperature from $10^{-12}$ s to geological time scales~\cite{Mydosh, Gatteschi03, Bedanta}.

In this Letter we report the observation of a new time-dependent magnetic behavior in which a transition from one long-range magnetically ordered state to another occurs in a zero magnetic field over a timescale of several hours! This phenomenon is revealed in a powder neutron-diffraction study of the Ising spin chain compound \caco. Our investigation focuses on the low-temperature regime ($T<14$~K) where our previous neutron-diffraction data have shown that there is an increasing instability in the spin density wave (SDW) order~\cite{Agrestini08b} within this material.
The results of the present work reveal that there is an order-order transition from the SDW structure to a commensurate antiferromagnetic (CAFM) phase. The time dependence of the magnetic reflections demonstrates that this transition occurs via a very slow transformation process. As the temperature is reduced the characteristic time of the transition process increases rapidly and at low-$T$ the magnetic states become frozen. 

\caco\ has a rhombohedral structure with spin chains made up of alternating face-sharing octahedral (Co$^{3+}_{\rm{I}};S=0$) and trigonal prismatic (Co$^{3+}_{\rm{II}};S=2$) CoO$_6$ polyhedra, running along the $c$ axis and arranged in a triangular lattice in the $ab$ plane~\cite{Fjellvag96}.
There is a strong intrachain ferromagnetic coupling ($J_{FM}\sim 25$~K)~\cite{Fresard04} and an Ising-like magnetic anisotropy with the easy direction parallel to the chains~\cite{Kageyama97,Wu05,Burnus06}.
A weaker AFM interchain coupling, ($J_{AFM}\sim 0.25$~K) stabilizes 3D incommensurate magnetic order at $T_N = 25$~K in the form of a longitudinal amplitude-modulated SDW propagating along the $c$ axis with a periodicity of $\sim 1000$~\AA. There is a phase shift of $120^\circ$ in the modulation between adjacent chains~\cite{Agrestini08,Agrestini08b}.
This magnetic structure can be modeled by considering two helical supersuperexchange pathways between adjacent spin chains, along with the intrachain FM exchange and a single ion anisotropy $DS^2$~\cite{Fresard04,Chapon09}. 

On further cooling the SDW become increasingly unstable and the volume of material exhibiting short-range order (SRO) increases~\cite{Agrestini08,Agrestini08b,Chapon09, Fleck10}. At 10 K the periodicity of the SDW is reported to slowly change with time~\cite{Moyoshi11}. A gradual slowing of the spin dynamics is reflected in a large frequency dependence of the ac susceptibility and a significant temperature hysteresis in the dc magnetization. At some spin-freeezing temperature, $T_{\rm{SF}}\sim7$~K, it is suggested that the spin relaxation time $\tau$ crosses from a $T$ dependent (Arrhenius) to a $T$ independent (quantum) regime~\cite{Hardy04b}. At low-$T$, in an applied magnetic field $H$, one observes a steplike behavior reminiscent of the quantum tunneling of magnetization, the appearance of a hysteresis in $M(H)$, and, on removal of the field, a slow relaxation in the magnetization~\cite{Kageyama97,Hardy03, Maignan04}.

A polycrystalline sample of \caco\ was synthesized via a solid-state method~\cite{Kageyama97}. The magnetic properties of the sample were checked by magnetization measurements and agree with those reported previously~\cite{Kageyama97,Hardy03}. Powder neutron-diffraction patterns were collected on the GEM time-of-flight instrument at the ISIS facility of the Rutherford Appleton Laboratory, UK. The diffraction data were refined using the FULLPROF program~\cite{Carvajal93, SuppNote}.
\begin{figure}[tb]
\begin{center}
\includegraphics[width=0.8\columnwidth]{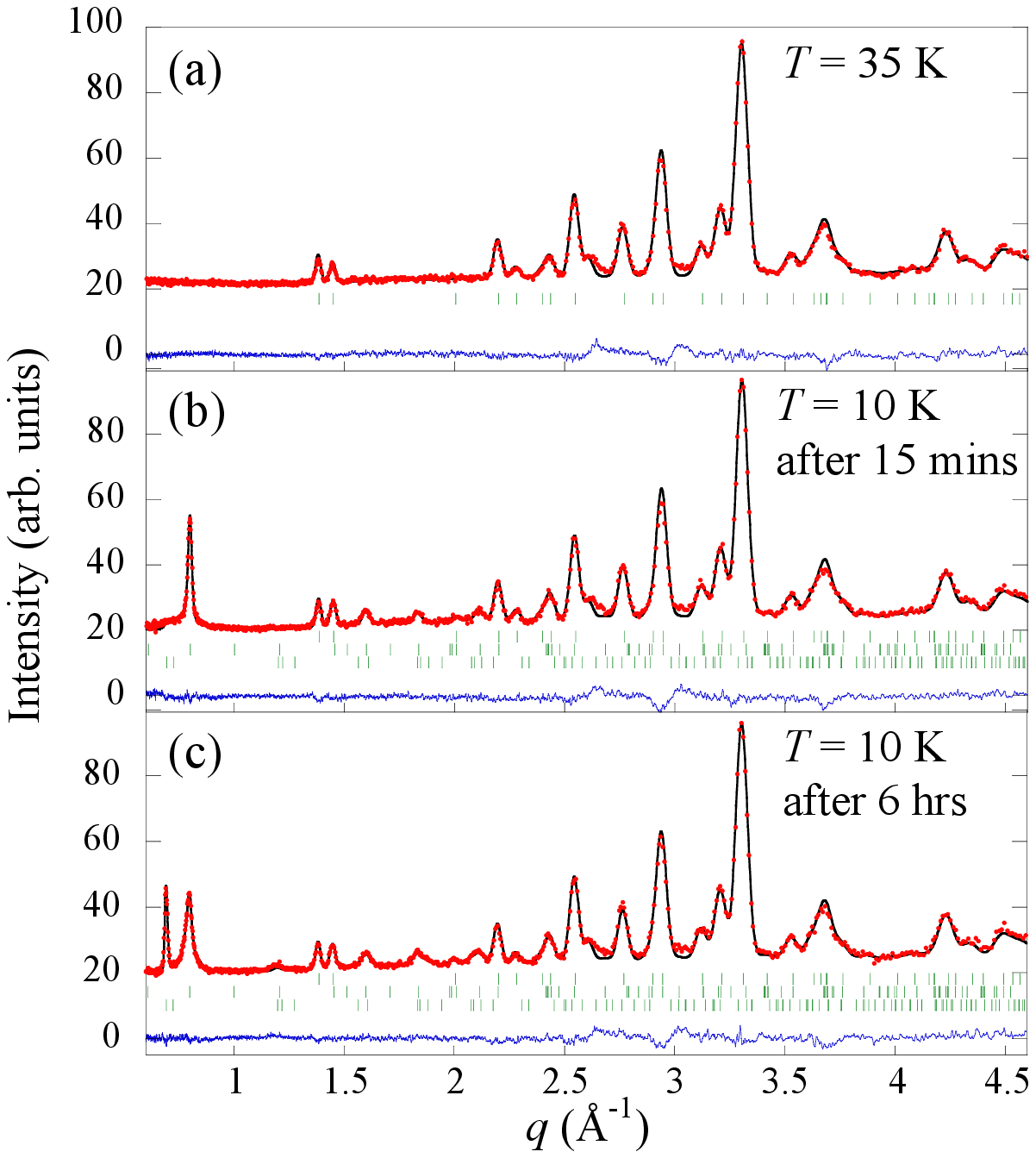}
\caption{\label{fig1} Powder neutron-diffraction patterns of \caco\ collected in 15 min at (a) 35~K ($T_N=25$~K) and (b) after waiting for 15~min and (c) 6~h at 10~K: experimental data (dots), calculated pattern (line) and difference curve. Ticks represent, from top to bottom, positions of the nuclear Bragg peaks and the magnetic peaks from the SDW and CAFM phases respectively.}
\end{center}
\end{figure}

An analysis of the neutron-diffraction data collected above $T_N$ at 35~K (Fig.~\ref{fig1}a) confirmed the high quality of our sample.
The results of the refinement of the nuclear structure are in good agreement with those in the literature~\cite{Aasland97,Fjellvag96} and show, to within the limits of the technique, that the sample is single phase.

\begin{figure}[tb]
\begin{center}
\includegraphics[width=0.8\columnwidth]{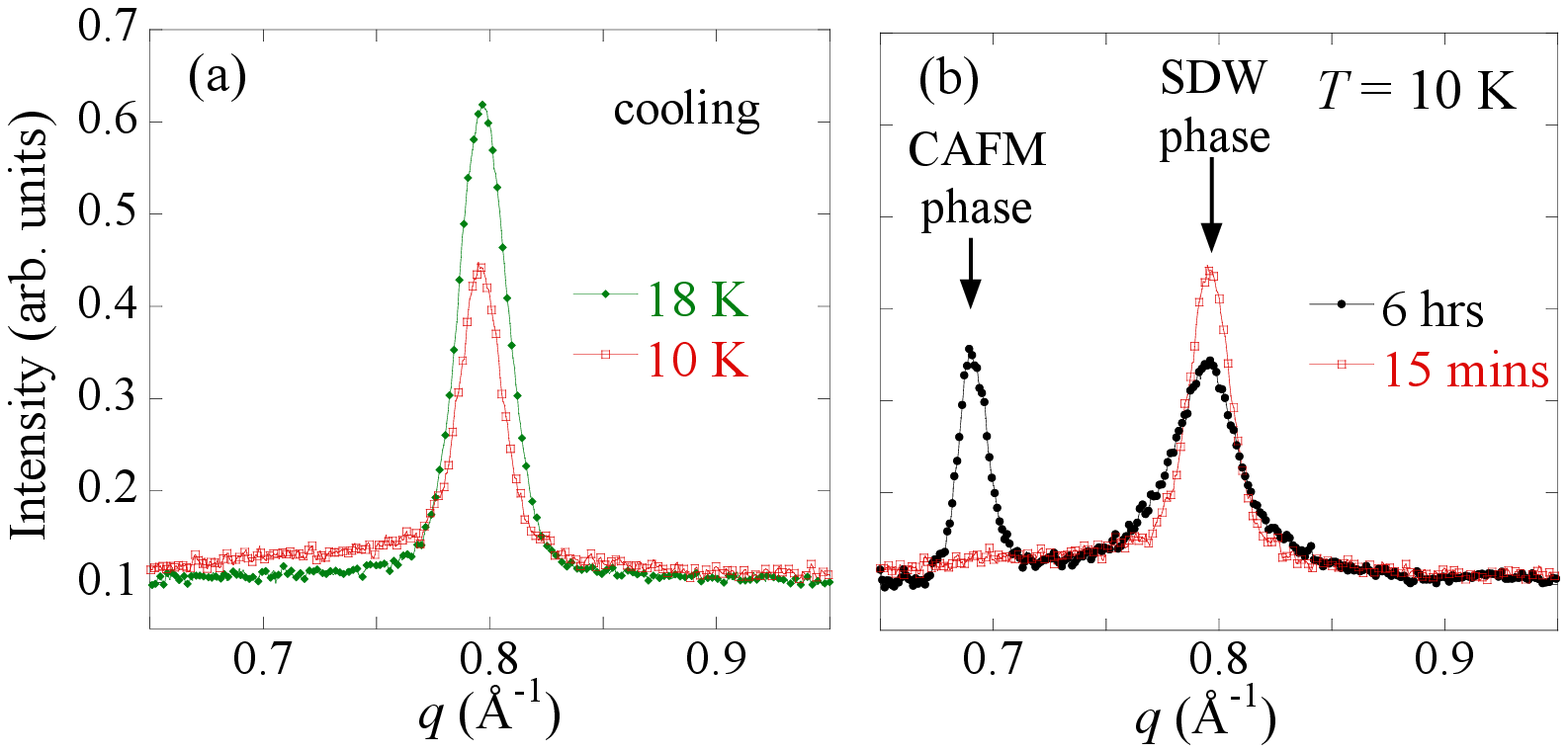}
\caption{\label{fig2}(color online). A selection of the results of the powder neutron-diffraction measurements on \caco. (a) Intensity of the $(1,0,0)$ magnetic Bragg peak of the SDW phase at different temperatures on cooling. (b) Neutron-diffraction patterns collected in 15 min at 10~K after different waiting times.
A new magnetic Bragg peak $(0.5,0.5,0)$ associated with the CAFM phase appears.}
\end{center}
\end{figure}

The antiferromagnetic reflections appearing in our data immediately  below $T_N$ can be indexed by the incommensurate propagation vector $\textbf{k}=(0,0,1.01)$ with respect to the hexagonal setting of the rhombohedral space group $R\overline{3}c$~\cite{SuppNote} and confirm that in this $T$ range the magnetic structure takes the form of a SDW~\cite{Agrestini08,Agrestini08b}. The integrated intensities of these reflections show a peculiar evolution as a function of temperature~\cite{Aasland97,Petrenko05,Agrestini08b,Fleck10}.
A broad maximum in the intensity of these peaks at 18~K is followed by a pronounced drop at lower temperatures. Our previous single-crystal neutron-diffraction studies on \caco\ have demonstrated that this is due, in part, to a reduction in the magnetic correlation lengths in the SDW phase~\cite{Agrestini08b,Fleck10}. The powder diffraction data of the present study confirm these findings. Fig.~\ref{fig2}a shows representative data collected over a limited $q$ range around the $\left(1, 0, 0\right)$ magnetic peak that characterizes the SDW phase. As the temperature is reduced the peaks associated with the SDW lose integrated intensity. Cooling to 10~K also results in the appearance of a plateau of intensity, not centered on the SDW magnetic Bragg peak and extending on the low $q$ side of the peak to $\sim$0.69~\AA$^{-1}$. We associate this previously unreported feature with the presence of a magnetic phase with short-range order. From this single feature we are unable to determine any details of the magnetic arrangement within this SRO phase.

\begin{figure}[tb]
\begin{center}
\includegraphics[width=0.8\columnwidth]{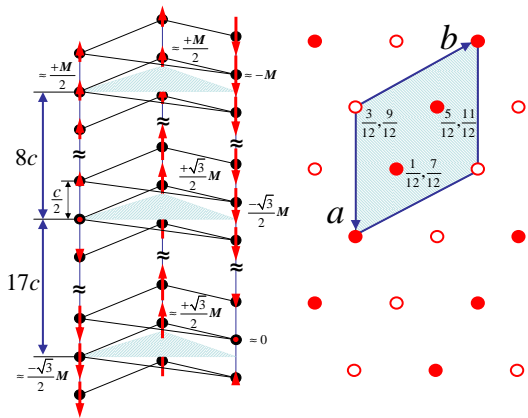}
\caption{\label{fig3}(color online). Schematics of the SDW and the CAFM phases. For the SDW phase, three sets of spin chains are shown, along with selected values for the magnetic moments on the trigonal prismatic Co$^{3+}_{\rm{II}}$ sites. The solid line indicates the dominant interchain, AFM supersuperexchange pathway~\cite{Fresard04,Agrestini08b, Chapon09}. In the CAFM phase the moments are aligned ferromagnetically along the $c$ axis respectively. The open and closed circles represent magnetic moments on the Co$^{3+}_{\rm{II}}$ sites pointing up and down the $c$ axis, along with the fractional coordinates of these sites along the $c$ axis within a unit cell.}
\end{center}
\end{figure}

\begin{figure}[tb]
\begin{center}
\includegraphics[width=0.8\columnwidth]{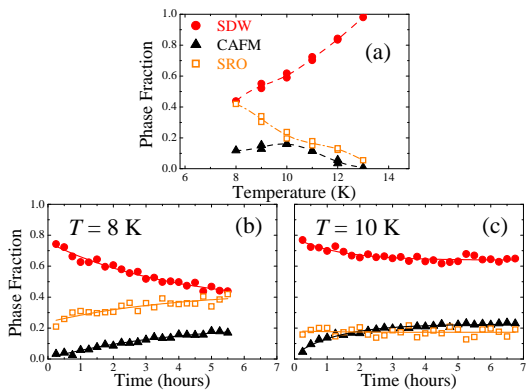}
\caption{\label{fig4}(color online). (a) Temperature evolution of the phase fractions of the SDW (closed circles), CAFM (closed triangles) and short-range (open squares) phases on warming after holding the sample at 8~K for 4 h. The lines are a guide to the eye.
(b), (c) Time dependence of the phase fractions of the SDW, CAFM and short-range phases at 8 and 10~K.
The curves are exponential fits to the data (see the text).}
\end{center}
\end{figure}

The most important result to emerge from the present work is the observation that if the sample is held at a constant temperature, ($8\leq T<12$~K), the magnetic reflections exhibit a clear \textit{time} dependence.  Fig.~\ref{fig2}b shows data collected at 10~K. All the magnetic reflections related to the SDW phase slowly lose intensity and become broader as a function of time. Simultaneously, new magnetic Bragg peaks begin to appear that after a few hours have become narrow and well defined.  These magnetic reflections can be indexed by the propagation vector $\textbf{k}=(0.5,-0.5,1)$ with respect to the hexagonal setting of the rhombohedral space group. Such a propagation vector indicates a doubling of the nuclear unit cell along both the $a$ and $b$ axes, which corresponds to an commensurate AFM arrangement of the spins in the $ab$ plane (see Fig.~\ref{fig3}). Along the $c$ axis the spins are ferromagnetically arranged confirming the strong FM intrachain coupling. An analysis of the intensity of the magnetic Bragg reflections of the CAFM phase shows that the moments are aligned along the $c$ axis. We note also that some of these new reflections [those with index $\left(h,-h, 0\right)$] are broadened, while others are resolution limited. This anisotropic line broadening reveals that the magnetic domains of the CAFM phase have a platelet shape, with a magnetic correlation length that is very short ($\sim$100~\AA) along the $[1,-1,0]$ direction. The interconversion of the magnetic phases is reversible. If the sample is warmed above 12~K the Bragg reflections of the CAFM phase disappear, while the reflections of the SDW structure again become narrow and intense. In Fig.~\ref{fig4}a we report the thermal evolution of the phase fractions on warming after waiting 4~h at 8~K. The vanishing of the CAFM phase at 12~K is clearly evident, while the volume of material exhibiting SRO is strongly reduced and the SDW phase is the majority phase.

In order to shed light on the unprecedented time dependence of magnetic order in \caco\ we collected neutron-diffraction data at fixed temperatures as a function of time. In Figs.~\ref{fig4}b and \ref{fig4}c we show the evolution of the phase fractions of the SDW, SRO, and CAFM phases at 8 and 10 K~\cite{SuppNote}. For each time dependence study the sample was first cooled rapidly ($\sim$2~K/min) from 18 to 2~K and then warmed to the desired temperature (8 or 10~K). The data were binned into time windows of 15 min. It is evident that a reduction of the volume of material exhibiting SDW order is mirrored by an increase in the amount of the CAFM and SRO phases present. The dynamics of this process are strongly temperature dependent. At 10~K an equilibrium regime is reached after 2 h, while at 8 K even after 5 h the system is still far from equilibrium. The diffraction patterns collected at 2~K showed no appreciable time dependence. If the sample is cooled rapidly from 18 to 2~K, the data contain only features corresponding to the SDW and the SRO phases. If the sample is cooled from 18 to 2~K, but held at 10~K for a time, the relative fractions of the three magnetic phases present at 10~K just before the final cooling step are frozen into the data at 2~K. This means that the characteristic time for phase interconversion increases rapidly as the temperature is reduced and at 2~K (data not shown) the process becomes so slow that it is impossible with neutron-diffraction to observe any change in the phase fractions over a practical timescale. 

We have fitted the phase fraction evolution by using a stretched exponential $\left(\beta=1\right)$ where the order parameter is the fraction of each of the three phases present. The relaxation time, $\tau$, which is the same for all three phases, increases from $\sim$$1.4\pm0.2$ to $\sim$$3.9\pm0.8$~h on cooling the sample from 10 to 8~K. At 8~K, the equilibrium fractions of the CAFM and SDW phases are approximately equal. The increase of the quantity of the CAFM phase present and the simultaneous reduction in the SDW phase fraction at equilibrium as the temperature is reduced indicate that the CAFM phase is the true ground state of \caco\ but that the two phases are nearly degenerate. An increase in the FWHM of the magnetic reflections characteristic of the SDW phase as $T$ is reduced corresponds to a reduction in the correlation length in the SDW phase from $\sim$500 to $\sim$300 \AA\, again confirming the instability of the SDW phase at low-$T$.

How can we explain the rich and complex magnetic behavior of \caco? Calculations show that the CAFM phase has a lower exchange energy than the SDW structure and is therefore expected to be the ground state of \caco\ ~\cite{Chapon09}. Our neutron data confirm the theoretical predictions, but also show that the SDW phase is preferred for 12~K~$< T <T_N$.
In order to explain this outcome we must take into account the entropy of the system. In the CAFM phase every magnetic site is fully ordered, while in the SDW phase there are regions in each chain with a nearly zero ordered moment. Thus, a difference in the configurational entropy of the two magnetic structures is sufficient, at higher temperatures, to stabilize the SDW phase~\cite{Chapon09}.

One key question concerns the mechanism underlying the slow interconversion of the two long-range ordered magnetic phases. In conventional magnetic systems a transition between two magnetically ordered phases is expected to occur instantaneously on the timescale of neutron diffraction~\cite{MotoyaNote}. Here, in contrast, the transition from the SDW to the CAFM phase occurs at a very slow rate. It is natural to suggest that the other unusual time-dependent magnetic behavior seen in \caco, such as the variation in the periodicity of the SDW, the slow relaxation of the magnetization, and the strong frequency dependence of the ac susceptibility, are all part of the same phenomenon. Drawing on an analogy with single molecule magnets, one can suppose that the strong intrachain FM coupling and the Ising-like magnetic anisotropy in \caco\ produce an energy barrier $\Delta$ between the spin-up and spin-down states which hampers the reversal of an individual spin. So, for example, the removal of an applied  magnetic field large enough to align all the spins within each chain results in a slow relaxation of the magnetization~\cite{Kageyama97,Hardy03, Maignan04}. Similar physics may also apply to the evolution of the 3D magnetic ordering in a zero magnetic field as the transition between the SDW and CAFM phases, that necessarily involves the production of phases with SRO, involves the flipping of a large number (theoretically an infinity) of spins in each chain. ac susceptibility measurements of the spin dynamics in \caco~\cite{Kageyama97,Hardy04b,Rayaprol03} revealed the presence of a thermally activated process. The relaxation time of this Arrhenius-like process increases rapidly as the $T$ is reduced until at $T\leq 8$~K it is supplanted by a $T$ independent (quantum) relaxation mechanism (see Fig. 4 of Ref.~\cite{Hardy04b}). The latter process may be associated with smaller spin units.

The variations in the widths of the magnetic peaks imply that the interconversion process proceeds via the intergrowth of the magnetic phases rather than by a mesoscopic phase separation. Although the CAFM phase is the ground state, at low $T$ it clearly does not percolate throughout the sample. Defects such as domain walls, that are already well established at higher $T$, could be one source of pinning that prevents the complete transformation to a single phase CAFM state.

In conclusion, we report experimental evidence of an ultraslow evolution between two long-range magnetically ordered states.
This work sheds new light on the nature of the magnetic order in \caco\ and the unusual loss of magnetic scattering intensity below 18~K in this material~\cite{Agrestini08b, Fleck10, Aasland97, Petrenko05}. It would be interesting to investigate whether similar slow order-order transitions are seen in other magnetic systems with degenerate ground states, or if this phenomenon is unique to \caco~\cite{MotoyaNote}.

This work was supported by the EPSRC, UK (EP/C000757/1) and by the European Commission under the 6th Framework Programme through the Key Action:Strengthening the European Research Area, Research Infrastructures. Contracts N$^\circ$:RII3-CT-2004-506008 and N$^\circ$~RII3-CT-2003-505925.


\begin{thebibliography}{100}
\bibitem{MotoyaNote} A recent neutron diffraction study of CeIr$_3$Si$_2$ revealed a transition between an incommensurate and a commensurate magnetic structure with similar dynamics~\cite{Motoya10}. In this case there is no change in the translational properties of the magnetic structure.
\end{thebibliography}
\end{document}


\title{Supplementary information}

\maketitle
\section{Experimental Methods}
A polycrystalline sample of \caco\ was synthesized via a solid-state method~\cite{Kageyama97}. Powder neutron-diffraction patterns were collected on the GEM time-of-flight instrument at the ISIS facility of the Rutherford Appleton Laboratory, UK. The diffraction data were refined using the FULLPROF program~\cite{Carvajal93}.

\section{Nuclear structure}

Table~\ref{Data35K} shows the results of the Rietveld refinement of the nuclear structure of Ca$_3$Co$_2$O$_6$ from powder neutron-diffraction data collected at 35 K. The profile function used was a pseudo-Voigt convoluted with Ikeda-Carpenter function.

\begin{table}[h]
	\centering
		\begin{tabular} {c c c c c c c c}\hline\hline 
		& Atom& Oxid. & Site & x& y & z & B$_{iso}$ (\AA$^2$) \\	\hline
		& Ca & +2 & 18e & 0.3688(3)& 0 & 1/4 & 0.07(2) \\
		& CoI & +3 & 6b & 0& 0 & 0 & 0.37(4) \\ 
		& CoII & +3 & 6a & 0& 0 & 1/4 & 0.15(4) \\ 
		& O & -2 & 36f & 0.1762(3) & 0.0245(2) & 0.1144(3) & 0.48(2) \\
			\hline\hline
		\end{tabular}
\caption{The atomic positions in Ca$_3$Co$_2$O$_6$ at 35 K. Space group: $R\bar{3}c$ (No. 167); $Z = 6$ in the hexagonal setting. Lattice parameters: $a = b = 9.0733(1)$~\AA, $c = 10.3830(3)$~\AA; $\alpha=90^\circ$, $\beta=90^\circ$,  $\gamma= 120^\circ$. $R_p=3.75$, $R_{wp}=5.06$, $\chi^2=3.71$.}
\label{Data35K}
\end{table}

\section{Magnetic structure}

From the results of previous neutron diffraction~\cite{Agrestini08b}, magnetic dichroism~\cite{Burnus06} and our magnetization measurements on single crystals of \caco, we estimate the magnetic moment on the high-spin ($S=2$) Co${\rm{II}}$ ion as $5.1 \pm 0.1 \mu_B$, including a sizable orbital moment. The magnetic moment of the Co${\rm{I}}$ ions is fixed at zero, because it is now well established~\cite{Sampa04,Burnus06} that the cobalt on the octahedral site is in a low-spin state (see Table~\ref{Data10K}). 

A Rietveld refinement of the powder neutron-diffraction data collected at 10 K shows that at this temperature the system contains three magnetic phases, a spin density wave (SDW) phase, a commensurate antiferromagnetic (CAFM) phase, and a phase exhibiting short-range order. The spin density wave (SDW) phase has a propagation vector $\textbf{k}=\left(0, 0, 1.01\right)$ while the commensurate antiferromagnetic (CAFM) phase with a propagation vector $\textbf{k}=\left(0.5, -0.5, 1\right)$~\cite{Note}.  The profile function used for the fitting was a pseudo-Voigt convoluted with Ikeda-Carpenter function. The fit made on the data collected after waiting for 15~minutes at 10~K gives $R_p=6.87$, $R_{wp}=8.67$, and $\chi^2=1.62$, while the fit made to the data collected after waiting 6~hours at 10~K gives $R_p=7.04$, $R_{wp}=8.84$, and $\chi^2=1.58$.

\begin{table}[h] 
	\centering
		\begin{tabular} {c c c c c c }\hline\hline
		& Atom & Site & M$_x$ & M$_y$ & M$_z$  \\	\hline 
		& CoI & 6b & 0 & 0 & 0  \\ 
		& CoII & 6a & 0 & 0 & 5.1  \\ 
			\hline\hline
		\end{tabular}
\caption{The magnitude and direction of the magnetic moments used in the Rietveld refinement of the magnetic structure. As we do not know the temperature dependence of the magnitude of the magnetic moments in each phase, it is problematic to distinguish between moment transfer and a change in the ratio of the phase fractions. In order to evaluate the variations in the phase fractions during the fitting, we have assumed that the moments are saturated. Given the low temperatures at which the fits are made and the fact that for all our fits the phase fractions sum to 1, we suggest that this assumption is reasonable.}
		\label{Data10K}
\end{table}

The anisotropic broadening of the CAFM phase was described by using the micro-structural model of platelet coherent domains, which assumes a Lorentzian contribution to the anisotropy being of the form $S_z\cdot\cos\phi$, where $S_z$ is the refined parameter and $\phi$ is the acute angle between the scattering vector $(h, k, l)$ and the vector $[1,-1, 0]$ defining the platelet shape of the domains. 
From the refined value $S_z = 10$ we have estimated a magnetic correlation length $D = C · K / S_z\approx100$~\AA, where $C$ is the diffractometer constant and $K$ is the Scherrer constant.

\section{Other Experiments}
The CAFM phase has also been observed in three other experiments, two on polycrystalline samples and one using a single crystal sample. The two neutron diffraction experiments on polycrystalline samples were carried out using the GEM instrument at ISIS and the D7 instrument at the Institute Laue Langevin (ILL). The results of the work on GEM are reported in Ref.~\cite{Agrestini08b}. The single crystal study was performed using I16 at the Diamond Light Source of the Rutherford Appleton Laboratory. No studies of the time dependence of the CAFM phase were performed during these experiments and none of the data collected are reported in this paper.